\documentclass[12pt]{article}                     
\usepackage{epic}
\newcommand{\Bbb}{\bf}

\def\ds{~\hbox{${\scriptstyle>\atop\raise3pt\hbox{$\scriptstyle<$}}$}~}

\def\hb#1{\hbox{#1}}

\def\+{\oplus}

\def\be{\begin{eqnarray}}
\def\ee{\end{eqnarray}}
\def\bea{\begin{eqnarray}}
\def\ena{\end{eqnarray}}
\def\bean{\begin{eqnarray*}}
\def\enan{\end{eqnarray*}}

\def\mref#1{(\ref{#1})}

\def\to{\longmapsto}

\def\C{{\bf C}}
\def\R{{\it R}}

\def\L{\Lambda}

\def\s{\sigma}
\def\z{\zeta}

\def\bra#1{\langle #1 |}       
\def\ket#1{| #1\rangle}         
\def\br#1{\langle #1 \rangle}   
\def\vac{|{\rm vac}\rangle}     
\def\dvac{\langle {\rm vac}|}   

\def\path{|p\rangle}

\def\End{{\rm End}\,}

\def\Tr{{\rm Tr}\,}

\def\com[#1,#2]{\hbox{$[#1,#2]$}}
\def\wt{{\rm wt}\,}
\def\uq{U_q(\widehat{sl}_2)}                             

\def\Phit{\widetilde{\Phi}}  
\def\Psit{\widetilde{\Psi}}  

\def\Phim#1{\mathrel{\mathop{\kern0pt \Phi}\limits^#1}}
\def\phim#1{\mathrel{\mathop{\kern0pt \phi}\limits^#1}}
\def\Psim#1{\mathrel{\mathop{\kern0pt \Psi}\limits^#1}}
\def\Phin#1{\mathrel{\mathop{\kern0pt \Phit}\limits^#1}}
\def\Psin#1{\mathrel{\mathop{\kern0pt \Psit}\limits^#1}}

\newcommand{\bqa}{\begin{eqnarray}}  
\newcommand{\eqa}{\end{eqnarray}}  
\newcommand{\ra}{\rightarrow}

\def\L{\Lambda}

\def\half{{1 \over 2}}  
\def\s{\sigma}

\def\d{\delta}

\def\ep{\epsilon} 
 
\def\l{\lambda}

\def\ov{\over}

\def\ed{\end{document}}
\def\ws{\;\;}

\def\ra{\rightarrow}  
  
\def\2pi{1\over 2\pi i}

\def\newline{\hfil\break}

\def\ra{\rightarrow}

\def\sq2{\sqrt{2}}  
\def\sqk2{\sqrt{2(k+2}}  
\def\sqk{\sqrt{k}}

\def\bfig{\begin{figure}}
\def\bfigt{\begin{figure}[top]}
\def\efig{\end{figure}}
\def\bea{\begin{eqnarray}}  
\def\eea{\end{eqnarray}}  
\def\br{\begin{array}}
\def\er{\end{array}}
\def\ea{\end{array}\end{equation}}
\def\eann{\end{array}\nonumber\end{equation}}
\def\bac{\begin{equation}\begin{array}{rll}}
\def\ba{\begin{equation}\begin{array}}
\def\nn{\nonumber} 
 
\def\wec#1#2#3#4#5{\overline{W}^1_{\ell}\left(\left.\begin{array}{ll}{#1}&{#2}\\{#3}&{#4}\end{array}\right|{#5}\right)}
\def\wo#1#2#3#4#5{W^I\left(\left.\begin{array}{ll}{#1}&{#2}\\{#3}&{#4}\end{array}\right|{#5}\right)}
\def\wt#1#2#3#4#5{W^{II}\left(\left.\begin{array}{ll}{#1}&{#2}\\{#3}&{#4}\end{array}\right|{#5}\right)}
\def\wth#1#2#3#4#5{W^*\left(\left.\begin{array}{ll}{#1}&{#2}\\{#3}&{#4}\end{array}\right|{#5}\right)}

\def\qp#1#2{({#1}\, ; \, {#2})_{\infty}}
\def\qp4#1{({#1}\, ; \, q^4)_{\infty}}

\def\br#1{({#1};\,x^{2r})_{\infty}}

\def\C{{\Bbb C}}

\def\cH{{\cal{H}}}

\def\cF{{\cal{F}}}
\def\cF{{\cal{F}}}

\def\cT{{\cal T}}

\def\sl{\sum\limits}

\def\ep{\varepsilon} 
\def\epp{\varepsilon^{\prime}}

\def\ot{\otimes}

\def\z{\zeta}

\def\bi{\bar{i}}
\def\bj{\bar{j}}
\def\bs{\tilde{s}}
\def\hf{{1\ov 2}}

\begin{document}
\begin{flushright}
hep-th/9804063 \\[10mm]
\end{flushright}
\begin{center}
{\Large \bf Mixing of Ground States in Vertex Models\\[8mm] }
{Jin Hong\,$^1$, Seok-Jin Kang\,$^1$, Tetsuji Miwa\,$^2$ and Robert Weston\,$^3$\\[8mm]
April 1998}
\end{center}
\footnotetext[1]{Dept of Mathematics, Seoul National
University, Seoul 151-742, Korea.}
\footnotetext[2]{Research Institute for Mathematical Sciences,
Kyoto University, Kyoto 606, Japan.}
\footnotetext[3]{Dept of Mathematics, Heriot-Watt University,
Edinburgh EH14 4AS, UK. }
\begin{abstract}
\noindent 
We consider the analogue of the 6-vertex model  constructed
from alternating spin $n/2$ and spin $m/2$ lines, where
$1\leq n<m$. We identify the transfer matrix and the space
on which it acts in terms of the representation theory of
$U_q(\widehat{sl}_2)$. We diagonalise the transfer matrix
and compute the S-matrix.
We give a trace formula for local correlation
functions. When $n=1$, the $1$-point function
of a spin $m/2$ local variable for the alternating lattice
with a particular ground state is given as a linear
combination of the $1$-point functions of the pure spin $m/2$
model with different ground states.  The mixing ratios are calculated
exactly and are expressed in terms of irreducible characters of
$U_q(\widehat{sl}_2)$ and the deformed Virasoro algebra.

\end{abstract}
An algebraic description of the infinite-volume 
six-vertex model is given in \cite{Daval,JM}.
There are two key elements in
this approach: to identify the half-infinite space
$\cdots \ot \C^2\ot \C^2 \ot\C^2 \ot \C^2$ on which 
the corner transfer matrix acts, 
with a level-one irreducible
highest-weight module
$V(\L_i)$ of $\uq$ (i=0,1) \cite{FM}; 
and to identify the half transfer matrices on this 
space with components of the $\uq$ intertwiner $V(\L_i) \ra
V(\L_{1-i})\ot \C^2$ \cite{JMN,JM}.

The six-vertex model is associated with a
lattice of intersecting lines, each carrying a spin $\half$
representation of $\uq$. In this letter, we consider the
analogous model on a lattice in which
spin $n/2$ and spin $m/2$ $(1\leq n<m)$ lines alternate.
Such models were constructed and analysed using the Bethe
Ansatz in \cite{VW92,VMN93,VMN94,AM93}.
The simplest example of the type of model we are considering
is the lattice consisting of alternating spin $\hf$ and spin $1$
lines, i.e., the case $n=1$, $m=2$ (see Figure 2).
The half-infinite space on which the corner transfer matrix
of this model acts is  $\cdots \ot \C^3\ot \C^2 \ot \C^3 \ot 
\C^2$. We shall develop an algebraic picture in which this
space is identified with 
the tensor 
product $\cF_{a,b}=V(\L_a)\ot V(\L_b)$, $(a,b=0,1)$. 
There will now be two half transfer matrices,
associated with spin $\hf$ and spin $1$ lines (see Figures 3a and 3b),
that we will identify with the $\uq$ intertwiners
\bea \phi^A&:&V(\L_a)\ot V(\L_b) \ra V(\L_a)\ot V(\L_{1-b})\ot
\C^{2}\ws \hb{and}\nn\\
\phi^B&:&V(\L_a)\ot V(\L_b) 
\ra V(\L_{1-a})\ot \C^2 \ot V(\L_b) \ra V(\L_{1-a})\ot V(\L_{1-b})\ot
\C^3\nn \ena
 respectively.
The full space $\cdots \C^{3}\ot \C^{2} \ot \C^{3} \ot 
\C^{2}\ot \C^{3}\ot \C^{2}\ot \cdots$ will then be identified with
the level zero module 
$\cF_{a,b}\ot \cF_{a,b}^*$.

We develop such a picture for the general case  $(1\leq n<m)$,
and construct 
the full transfer-matrix in terms of the 
intertwiners $\phi^A$ and $\phi^B$ and their duals.
We diagonalise the transfer matrix, and find a pseudo-particle spectrum
consisting of only spin $0$ and
spin $\half$ states.  Such a spectrum was observed in the case
$n=1$, $m=2$ in \cite{VMN94}. 
We compute the S-matrix for these excitations,
and give a trace
formula for local correlation functions of the theory.

Using our algebraic picture,
we show that the $1$-point function of a spin $n/2$
variable is equal to that in the pure spin $n/2$ lattice when the ground states
of the two different lattices are chosen correctly. However, we also show that
the $1$-point function of a spin $m/2$ variable
is a linear combination of the ones for the pure spin $m/2$ lattice
corresponding to different ground states. We call this result
a {\it mixing of ground states}.


After this brief introduction to some of our results, we
will now define our alternating spin vertex
model in some more detail.
First of all,
the Boltzmann weights are specified by the \R-matrix 
$R^{(k,l)}(\z_1/\z_2):V^{(k)}_{\z_1} \ot
V^{(l)}_{\z_2} \ra V^{(k)}_{\z_1} \ot V^{(l)}_{\z_2}$,
where $V^{(k)}_{\z}\simeq \C^{k+1}$ is the $\uq$ principal evaluation module,
with weight vectors $u^{(k)}_i$ $(i=0,\cdots,k)$, given in \cite{MW97}.
We fix the normalisation such that 
$R^{(k,l)}(\z)=\bar{R}^{(k,l)}(\z)/\kappa^{(k,l)}(\z)$, where
$\bar{R}^{(k,l)}(\z)(u^{(k)}_0 \ot u^{(l)}_0)=(u^{(k)}_0 \ot
u^{(l)}_0)$,
and
\be \kappa^{(k,l)}(\z)=\z^{min(k,l)} \frac{\qp4{q^{2+k+l}\z^2}
  \qp4{q^{2+|k-l|}\z^{-2}}}%
{\qp4{q^{2+k+l}\z^{-2}} \qp4{q^{2+|k-l|}\z^2} }.\nn\ee
With this normalisation, the partition function 
per unit site of our lattice model will be 1.

A lattice vertex associated with the intersection of a spin $k/2$ and
spin $l/2$ line has 6 variables attached to it: spin variables
$i,i'=(0,\cdots,k)$ and
$j,j'=(0,\cdots,l)$, and  spectral parameters 
$\z_1,\z_2\in \C$. A Boltzmann weight 
$R^{(k,l)}(\z)^{i,j}_{i',j'}$ (with $\z=\z_1/\z_2$) is attached to the following
configuration of these variables:

\setlength{\unitlength}{1.2pt}%
 \begin{picture}(100,80)(-100,13)
\thicklines
\put(50,75){\vector(0,-1){50}}
\put(75,50){\vector(-1,0){50}}
\put(50,78){\makebox(0,0)[b]{$i$}}
\put(78,50){\makebox(0,0)[l]{$j$}}
\put(50,22){\makebox(0,0)[t]{$i'$}}
\put(22,50){\makebox(0,0)[r]{$j'$}}
\put(49,63){\makebox(0,0)[r]{$\z_1$}}
\put(63,48){\makebox(0,0)[t]{$\z_2$}}
\put(-100,45){\makebox(0,0)[lb]{Figure 1}}

\end{picture}

We choose to restrict our parameters to the region $-1<q<0$, 
$1<\z <(-q)^{-1}$. With this restriction,  the weights 
$R^{(k,l)}(\z)^{i,j}_{i',j'}$ with $k\leq l$ 
that are of the lowest order, 
and are larger than all other weights of higher order,
are all those that obey the requirement $k\leq i+j \leq l$, $i'=k-i$,
$j'=j+2i-k$. For example, if $k=1$ and $l=2$, the largest
Boltzmann weights will be $R^{(1,2)}(\z)^{0,1}_{1,0}$,
$R^{(1,2)}(\z)^{0,2}_{1,1}$, $R^{(1,2)}(\z)^{1,0}_{0,1}$, and
$R^{(1,2)}(\z)^{1,1}_{0,2}$.

Now consider the finite periodic lattice constructed by 
alternating 
$2M$ spin $n/2$ with $2M$ spin $m/2$ lines in both the horizontal 
and vertical directions. Such a lattice with $M=2$ is shown in
Figure 2. In this figure, solid lines represent spin $n/2$ lines,
and dotted lines spin $m/2$ lines. The vertical lines have spectral
parameter $\z$, and horizontal lines have a spectral paramater equal to $1$.
We place two horizontal and two vertical 
spin $n/2$ lines next to each other 
at the centre of the lattice, and two vertical and two horizontal
spin $m/2$ lines lines next
to each other at the boundaries. We do this purely to simplify
the discussion of the spaces on which the
corner transfer matrices act. The order of the spin $n/2$ and spin $m/2$ lines
can be switched to a uniform alternating pattern by using the Yang-Baxter
equation.

\setlength{\unitlength}{0.00067in}
\begingroup\makeatletter\ifx\SetFigFont\undefined%
\gdef\SetFigFont#1#2#3#4#5{%
  \reset@font\fontsize{#1}{#2pt}%
  \fontfamily{#3}\fontseries{#4}\fontshape{#5}%
  \selectfont}%
\fi\endgroup%
{\renewcommand{\dashlinestretch}{30}
\begin{picture}(7437,6500)(0,-350)
\drawline(4350,5412)(4350,1212)
\drawline(4950,5412)(4950,1212)
\drawline(4950,1212)(4950,12)
\drawline(4920.000,132.000)(4950.000,12.000)(4980.000,132.000)
\drawline(4350,1212)(4350,12)
\drawline(4320.000,132.000)(4350.000,12.000)(4380.000,132.000)
\drawline(6150,5412)(6150,12)
\drawline(6120.000,132.000)(6150.000,12.000)(6180.000,132.000)
\drawline(3150,5412)(3150,12)
\drawline(3120.000,132.000)(3150.000,12.000)(3180.000,132.000)
\drawline(1950,4212)(7350,4212)
\drawline(2070.000,4242.000)(1950.000,4212.000)(2070.000,4182.000)
\drawline(1950,3012)(7350,3012)
\drawline(2070.000,3042.000)(1950.000,3012.000)(2070.000,2982.000)
\drawline(1950,2412)(7350,2412)
\drawline(2070.000,2442.000)(1950.000,2412.000)(2070.000,2382.000)
\drawline(1950,1212)(7350,1212)
\drawline(2070.000,1242.000)(1950.000,1212.000)(2070.000,1182.000)
\dashline{60.000}(5550,5412)(5550,12)
\drawline(5520.000,132.000)(5550.000,12.000)(5580.000,132.000)
\dashline{60.000}(6750,5412)(6750,12)
\drawline(6720.000,132.000)(6750.000,12.000)(6780.000,132.000)
\dashline{60.000}(3750,5412)(3750,12)
\drawline(3720.000,132.000)(3750.000,12.000)(3780.000,132.000)
\dashline{60.000}(2550,5412)(2550,12)
\drawline(2520.000,132.000)(2550.000,12.000)(2580.000,132.000)
\dashline{60.000}(1950,612)(7350,612)
\drawline(2070.000,642.000)(1950.000,612.000)(2070.000,582.000)
\dashline{60.000}(1950,1812)(7350,1812)
\drawline(2070.000,1842.000)(1950.000,1812.000)(2070.000,1782.000)
\dashline{60.000}(1950,3612)(7425,3612)
\drawline(2070.000,3642.000)(1950.000,3612.000)(2070.000,3582.000)
\dashline{60.000}(1950,4812)(7350,4812)
\drawline(2070.000,4842.000)(1950.000,4812.000)(2070.000,4782.000)
\put(6975,387){\makebox(0,0)[lb]{\smash{{{\SetFigFont{8}{14.4}{\rmdefault}{\mddefault}{\updefault}$\bar{j_1}$}}}}}
\put(6975,987){\makebox(0,0)[lb]{\smash{{{\SetFigFont{8}{14.4}{\rmdefault}{\mddefault}{\updefault}$\bar{i}$}}}}}
\put(6975,1587){\makebox(0,0)[lb]{\smash{{{\SetFigFont{8}{14.4}{\rmdefault}{\mddefault}{\updefault}$\bar{j}$}}}}}
\put(6975,2187){\makebox(0,0)[lb]{\smash{{{\SetFigFont{8}{14.4}{\rmdefault}{\mddefault}{\updefault}$\bar{i}$}}}}}
\put(6975,2787){\makebox(0,0)[lb]{\smash{{{\SetFigFont{8}{14.4}{\rmdefault}{\mddefault}{\updefault}$i$}}}}}
\put(6975,3387){\makebox(0,0)[lb]{\smash{{{\SetFigFont{8}{14.4}{\rmdefault}{\mddefault}{\updefault}$j$}}}}}
\put(6975,3987){\makebox(0,0)[lb]{\smash{{{\SetFigFont{8}{14.4}{\rmdefault}{\mddefault}{\updefault}$i$}}}}}
\put(6975,4587){\makebox(0,0)[lb]{\smash{{{\SetFigFont{8}{14.4}{\rmdefault}{\mddefault}{\updefault}$j_1$}}}}}
\put(6375,387){\makebox(0,0)[lb]{\smash{{{\SetFigFont{8}{14.4}{\rmdefault}{\mddefault}{\updefault}$j_1$}}}}}
\put(6375,987){\makebox(0,0)[lb]{\smash{{{\SetFigFont{8}{14.4}{\rmdefault}{\mddefault}{\updefault}$i$}}}}}
\put(6375,1587){\makebox(0,0)[lb]{\smash{{{\SetFigFont{8}{14.4}{\rmdefault}{\mddefault}{\updefault}$j$}}}}}
\put(6375,2187){\makebox(0,0)[lb]{\smash{{{\SetFigFont{8}{14.4}{\rmdefault}{\mddefault}{\updefault}$i$}}}}}
\put(6375,2787){\makebox(0,0)[lb]{\smash{{{\SetFigFont{8}{14.4}{\rmdefault}{\mddefault}{\updefault}$\bi$}}}}}
\put(6375,3387){\makebox(0,0)[lb]{\smash{{{\SetFigFont{8}{14.4}{\rmdefault}{\mddefault}{\updefault}$\bj$}}}}}
\put(6375,3987){\makebox(0,0)[lb]{\smash{{{\SetFigFont{8}{14.4}{\rmdefault}{\mddefault}{\updefault}$\bi$}}}}}
\put(6375,4587){\makebox(0,0)[lb]{\smash{{{\SetFigFont{8}{14.4}{\rmdefault}{\mddefault}{\updefault}$\bj_1$}}}}}
\put(5775,387){\makebox(0,0)[lb]{\smash{{{\SetFigFont{8}{14.4}{\rmdefault}{\mddefault}{\updefault}$\bar{j}$}}}}}
\put(5775,987){\makebox(0,0)[lb]{\smash{{{\SetFigFont{8}{14.4}{\rmdefault}{\mddefault}{\updefault}$\bar{i}$}}}}}
\put(5775,1587){\makebox(0,0)[lb]{\smash{{{\SetFigFont{8}{14.4}{\rmdefault}{\mddefault}{\updefault}$\bar{j_1}$}}}}}
\put(5775,2187){\makebox(0,0)[lb]{\smash{{{\SetFigFont{8}{14.4}{\rmdefault}{\mddefault}{\updefault}$\bar{i}$}}}}}
\put(5775,2787){\makebox(0,0)[lb]{\smash{{{\SetFigFont{8}{14.4}{\rmdefault}{\mddefault}{\updefault}$i$}}}}}
\put(5775,3387){\makebox(0,0)[lb]{\smash{{{\SetFigFont{8}{14.4}{\rmdefault}{\mddefault}{\updefault}$j_1$}}}}}
\put(5775,3987){\makebox(0,0)[lb]{\smash{{{\SetFigFont{8}{14.4}{\rmdefault}{\mddefault}{\updefault}$i$}}}}}
\put(5775,4587){\makebox(0,0)[lb]{\smash{{{\SetFigFont{8}{14.4}{\rmdefault}{\mddefault}{\updefault}$j$}}}}}
\put(5175,387){\makebox(0,0)[lb]{\smash{{{\SetFigFont{8}{14.4}{\rmdefault}{\mddefault}{\updefault}$j$}}}}}
\put(5175,987){\makebox(0,0)[lb]{\smash{{{\SetFigFont{8}{14.4}{\rmdefault}{\mddefault}{\updefault}$i$}}}}}
\put(5175,1587){\makebox(0,0)[lb]{\smash{{{\SetFigFont{8}{14.4}{\rmdefault}{\mddefault}{\updefault}$j_1$}}}}}
\put(5175,2187){\makebox(0,0)[lb]{\smash{{{\SetFigFont{8}{14.4}{\rmdefault}{\mddefault}{\updefault}$i$}}}}}
\put(5175,2787){\makebox(0,0)[lb]{\smash{{{\SetFigFont{8}{14.4}{\rmdefault}{\mddefault}{\updefault}$\bi$}}}}}
\put(5175,3387){\makebox(0,0)[lb]{\smash{{{\SetFigFont{8}{14.4}{\rmdefault}{\mddefault}{\updefault}$\bj_1$}}}}}
\put(5175,3987){\makebox(0,0)[lb]{\smash{{{\SetFigFont{8}{14.4}{\rmdefault}{\mddefault}{\updefault}$\bi$}}}}}
\put(5175,4587){\makebox(0,0)[lb]{\smash{{{\SetFigFont{8}{14.4}{\rmdefault}{\mddefault}{\updefault}$\bj$}}}}}
\put(4575,387){\makebox(0,0)[lb]{\smash{{{\SetFigFont{8}{14.4}{\rmdefault}{\mddefault}{\updefault}$\bar{j_1}$}}}}}
\put(4575,987){\makebox(0,0)[lb]{\smash{{{\SetFigFont{8}{14.4}{\rmdefault}{\mddefault}{\updefault}$\bar{i}$}}}}}
\put(4575,1587){\makebox(0,0)[lb]{\smash{{{\SetFigFont{8}{14.4}{\rmdefault}{\mddefault}{\updefault}$\bar{j}$}}}}}
\put(4575,2187){\makebox(0,0)[lb]{\smash{{{\SetFigFont{8}{14.4}{\rmdefault}{\mddefault}{\updefault}$\bar{i}$}}}}}
\put(4575,2787){\makebox(0,0)[lb]{\smash{{{\SetFigFont{8}{14.4}{\rmdefault}{\mddefault}{\updefault}$i$}}}}}
\put(4575,3387){\makebox(0,0)[lb]{\smash{{{\SetFigFont{8}{14.4}{\rmdefault}{\mddefault}{\updefault}$j$}}}}}
\put(4575,3987){\makebox(0,0)[lb]{\smash{{{\SetFigFont{8}{14.4}{\rmdefault}{\mddefault}{\updefault}$i$}}}}}
\put(4575,4587){\makebox(0,0)[lb]{\smash{{{\SetFigFont{8}{14.4}{\rmdefault}{\mddefault}{\updefault}$j_1$}}}}}
\put(3975,387){\makebox(0,0)[lb]{\smash{{{\SetFigFont{8}{14.4}{\rmdefault}{\mddefault}{\updefault}$j$}}}}}
\put(3975,987){\makebox(0,0)[lb]{\smash{{{\SetFigFont{8}{14.4}{\rmdefault}{\mddefault}{\updefault}$i$}}}}}
\put(3975,1587){\makebox(0,0)[lb]{\smash{{{\SetFigFont{8}{14.4}{\rmdefault}{\mddefault}{\updefault}$j_1$}}}}}
\put(3975,2187){\makebox(0,0)[lb]{\smash{{{\SetFigFont{8}{14.4}{\rmdefault}{\mddefault}{\updefault}$i$}}}}}
\put(3975,2787){\makebox(0,0)[lb]{\smash{{{\SetFigFont{8}{14.4}{\rmdefault}{\mddefault}{\updefault}$\bi$}}}}}
\put(3975,3387){\makebox(0,0)[lb]{\smash{{{\SetFigFont{8}{14.4}{\rmdefault}{\mddefault}{\updefault}$\bj_1$}}}}}
\put(3975,3987){\makebox(0,0)[lb]{\smash{{{\SetFigFont{8}{14.4}{\rmdefault}{\mddefault}{\updefault}$\bi$}}}}}
\put(3975,4587){\makebox(0,0)[lb]{\smash{{{\SetFigFont{8}{14.4}{\rmdefault}{\mddefault}{\updefault}$\bj$}}}}}
\put(3375,387){\makebox(0,0)[lb]{\smash{{{\SetFigFont{8}{14.4}{\rmdefault}{\mddefault}{\updefault}$\bar{j}$}}}}}
\put(3375,987){\makebox(0,0)[lb]{\smash{{{\SetFigFont{8}{14.4}{\rmdefault}{\mddefault}{\updefault}$\bar{i}$}}}}}
\put(3375,1587){\makebox(0,0)[lb]{\smash{{{\SetFigFont{8}{14.4}{\rmdefault}{\mddefault}{\updefault}$\bar{j_1}$}}}}}
\put(3375,2187){\makebox(0,0)[lb]{\smash{{{\SetFigFont{8}{14.4}{\rmdefault}{\mddefault}{\updefault}$\bar{i}$}}}}}
\put(3375,2787){\makebox(0,0)[lb]{\smash{{{\SetFigFont{8}{14.4}{\rmdefault}{\mddefault}{\updefault}$i$}}}}}
\put(3375,3387){\makebox(0,0)[lb]{\smash{{{\SetFigFont{8}{14.4}{\rmdefault}{\mddefault}{\updefault}$j_1$}}}}}
\put(3375,3987){\makebox(0,0)[lb]{\smash{{{\SetFigFont{8}{14.4}{\rmdefault}{\mddefault}{\updefault}$i$}}}}}
\put(3375,4587){\makebox(0,0)[lb]{\smash{{{\SetFigFont{8}{14.4}{\rmdefault}{\mddefault}{\updefault}$j$}}}}}
\put(2775,387){\makebox(0,0)[lb]{\smash{{{\SetFigFont{8}{14.4}{\rmdefault}{\mddefault}{\updefault}$j_1$}}}}}
\put(2775,987){\makebox(0,0)[lb]{\smash{{{\SetFigFont{8}{14.4}{\rmdefault}{\mddefault}{\updefault}$i$}}}}}
\put(2775,1587){\makebox(0,0)[lb]{\smash{{{\SetFigFont{8}{14.4}{\rmdefault}{\mddefault}{\updefault}$j$}}}}}
\put(2775,2187){\makebox(0,0)[lb]{\smash{{{\SetFigFont{8}{14.4}{\rmdefault}{\mddefault}{\updefault}$i$}}}}}
\put(2775,2787){\makebox(0,0)[lb]{\smash{{{\SetFigFont{8}{14.4}{\rmdefault}{\mddefault}{\updefault}$\bi$}}}}}
\put(2775,3387){\makebox(0,0)[lb]{\smash{{{\SetFigFont{8}{14.4}{\rmdefault}{\mddefault}{\updefault}$\bj$}}}}}
\put(2775,3987){\makebox(0,0)[lb]{\smash{{{\SetFigFont{8}{14.4}{\rmdefault}{\mddefault}{\updefault}$\bi$}}}}}
\put(2775,4587){\makebox(0,0)[lb]{\smash{{{\SetFigFont{8}{14.4}{\rmdefault}{\mddefault}{\updefault}$\bj_1$}}}}}
\put(2175,387){\makebox(0,0)[lb]{\smash{{{\SetFigFont{8}{14.4}{\rmdefault}{\mddefault}{\updefault}$\bar{j_1}$}}}}}
\put(2175,987){\makebox(0,0)[lb]{\smash{{{\SetFigFont{8}{14.4}{\rmdefault}{\mddefault}{\updefault}$\bar{i}$}}}}}
\put(2175,1587){\makebox(0,0)[lb]{\smash{{{\SetFigFont{8}{14.4}{\rmdefault}{\mddefault}{\updefault}$\bar{j}$}}}}}
\put(2175,2187){\makebox(0,0)[lb]{\smash{{{\SetFigFont{8}{14.4}{\rmdefault}{\mddefault}{\updefault}$\bar{i}$}}}}}
\put(2175,2787){\makebox(0,0)[lb]{\smash{{{\SetFigFont{8}{14.4}{\rmdefault}{\mddefault}{\updefault}$i$}}}}}
\put(2175,3387){\makebox(0,0)[lb]{\smash{{{\SetFigFont{8}{14.4}{\rmdefault}{\mddefault}{\updefault}$j$}}}}}
\put(2175,3987){\makebox(0,0)[lb]{\smash{{{\SetFigFont{8}{14.4}{\rmdefault}{\mddefault}{\updefault}$i$}}}}}
\put(2175,4587){\makebox(0,0)[lb]{\smash{{{\SetFigFont{8}{14.4}{\rmdefault}{\mddefault}{\updefault}$j_1$}}}}}
\put(2400,5037){\makebox(0,0)[lb]{\smash{{{\SetFigFont{8}{14.4}{\rmdefault}{\mddefault}{\updefault}$j_1$}}}}}
\put(2400,4437){\makebox(0,0)[lb]{\smash{{{\SetFigFont{8}{14.4}{\rmdefault}{\mddefault}{\updefault}$\bj_1$}}}}}
\put(2400,3837){\makebox(0,0)[lb]{\smash{{{\SetFigFont{8}{14.4}{\rmdefault}{\mddefault}{\updefault}$j$}}}}}
\put(2400,3237){\makebox(0,0)[lb]{\smash{{{\SetFigFont{8}{14.4}{\rmdefault}{\mddefault}{\updefault}$\bj$}}}}}
\put(2400,2637){\makebox(0,0)[lb]{\smash{{{\SetFigFont{8}{14.4}{\rmdefault}{\mddefault}{\updefault}$j_1$}}}}}
\put(2400,2037){\makebox(0,0)[lb]{\smash{{{\SetFigFont{8}{14.4}{\rmdefault}{\mddefault}{\updefault}$\bj$}}}}}
\put(2400,1437){\makebox(0,0)[lb]{\smash{{{\SetFigFont{8}{14.4}{\rmdefault}{\mddefault}{\updefault}$j$}}}}}
\put(2400,837){\makebox(0,0)[lb]{\smash{{{\SetFigFont{8}{14.4}{\rmdefault}{\mddefault}{\updefault}$\bj_1$}}}}}
\put(2400,237){\makebox(0,0)[lb]{\smash{{{\SetFigFont{8}{14.4}{\rmdefault}{\mddefault}{\updefault}$j_1$}}}}}
\put(3000,5037){\makebox(0,0)[lb]{\smash{{{\SetFigFont{8}{14.4}{\rmdefault}{\mddefault}{\updefault}$i$}}}}}
\put(3000,4437){\makebox(0,0)[lb]{\smash{{{\SetFigFont{8}{14.4}{\rmdefault}{\mddefault}{\updefault}$\bi$}}}}}
\put(3000,3837){\makebox(0,0)[lb]{\smash{{{\SetFigFont{8}{14.4}{\rmdefault}{\mddefault}{\updefault}$i$}}}}}
\put(3000,3237){\makebox(0,0)[lb]{\smash{{{\SetFigFont{8}{14.4}{\rmdefault}{\mddefault}{\updefault}$\bi$}}}}}
\put(3000,2637){\makebox(0,0)[lb]{\smash{{{\SetFigFont{8}{14.4}{\rmdefault}{\mddefault}{\updefault}$i$}}}}}
\put(3000,2037){\makebox(0,0)[lb]{\smash{{{\SetFigFont{8}{14.4}{\rmdefault}{\mddefault}{\updefault}$\bi$}}}}}
\put(3000,1437){\makebox(0,0)[lb]{\smash{{{\SetFigFont{8}{14.4}{\rmdefault}{\mddefault}{\updefault}$i$}}}}}
\put(3000,837){\makebox(0,0)[lb]{\smash{{{\SetFigFont{8}{14.4}{\rmdefault}{\mddefault}{\updefault}$\bi$}}}}}
\put(3000,237){\makebox(0,0)[lb]{\smash{{{\SetFigFont{8}{14.4}{\rmdefault}{\mddefault}{\updefault}$i$}}}}}
\put(3600,5037){\makebox(0,0)[lb]{\smash{{{\SetFigFont{8}{14.4}{\rmdefault}{\mddefault}{\updefault}$j$}}}}}
\put(3600,4437){\makebox(0,0)[lb]{\smash{{{\SetFigFont{8}{14.4}{\rmdefault}{\mddefault}{\updefault}$\bj$}}}}}
\put(3600,3837){\makebox(0,0)[lb]{\smash{{{\SetFigFont{8}{14.4}{\rmdefault}{\mddefault}{\updefault}$j_1$}}}}}
\put(3600,3237){\makebox(0,0)[lb]{\smash{{{\SetFigFont{8}{14.4}{\rmdefault}{\mddefault}{\updefault}$\bj_1$}}}}}
\put(3600,2637){\makebox(0,0)[lb]{\smash{{{\SetFigFont{8}{14.4}{\rmdefault}{\mddefault}{\updefault}$j$}}}}}
\put(3600,2037){\makebox(0,0)[lb]{\smash{{{\SetFigFont{8}{14.4}{\rmdefault}{\mddefault}{\updefault}$\bj_1$}}}}}
\put(3600,1437){\makebox(0,0)[lb]{\smash{{{\SetFigFont{8}{14.4}{\rmdefault}{\mddefault}{\updefault}$j_1$}}}}}
\put(3600,837){\makebox(0,0)[lb]{\smash{{{\SetFigFont{8}{14.4}{\rmdefault}{\mddefault}{\updefault}$\bj$}}}}}
\put(3600,237){\makebox(0,0)[lb]{\smash{{{\SetFigFont{8}{14.4}{\rmdefault}{\mddefault}{\updefault}$j$}}}}}
\put(4200,5037){\makebox(0,0)[lb]{\smash{{{\SetFigFont{8}{14.4}{\rmdefault}{\mddefault}{\updefault}$i$}}}}}
\put(4200,4437){\makebox(0,0)[lb]{\smash{{{\SetFigFont{8}{14.4}{\rmdefault}{\mddefault}{\updefault}$\bi$}}}}}
\put(4200,3837){\makebox(0,0)[lb]{\smash{{{\SetFigFont{8}{14.4}{\rmdefault}{\mddefault}{\updefault}$i$}}}}}
\put(4200,3237){\makebox(0,0)[lb]{\smash{{{\SetFigFont{8}{14.4}{\rmdefault}{\mddefault}{\updefault}$\bi$}}}}}
\put(4200,2637){\makebox(0,0)[lb]{\smash{{{\SetFigFont{8}{14.4}{\rmdefault}{\mddefault}{\updefault}$i$}}}}}
\put(4200,2037){\makebox(0,0)[lb]{\smash{{{\SetFigFont{8}{14.4}{\rmdefault}{\mddefault}{\updefault}$\bi$}}}}}
\put(4200,1437){\makebox(0,0)[lb]{\smash{{{\SetFigFont{8}{14.4}{\rmdefault}{\mddefault}{\updefault}$i$}}}}}
\put(4200,837){\makebox(0,0)[lb]{\smash{{{\SetFigFont{8}{14.4}{\rmdefault}{\mddefault}{\updefault}$\bi$}}}}}
\put(4200,237){\makebox(0,0)[lb]{\smash{{{\SetFigFont{8}{14.4}{\rmdefault}{\mddefault}{\updefault}$i$}}}}}
\put(4800,5037){\makebox(0,0)[lb]{\smash{{{\SetFigFont{8}{14.4}{\rmdefault}{\mddefault}{\updefault}$\bi$}}}}}
\put(4800,4437){\makebox(0,0)[lb]{\smash{{{\SetFigFont{8}{14.4}{\rmdefault}{\mddefault}{\updefault}$i$}}}}}
\put(4800,3837){\makebox(0,0)[lb]{\smash{{{\SetFigFont{8}{14.4}{\rmdefault}{\mddefault}{\updefault}$\bi$}}}}}
\put(4800,3237){\makebox(0,0)[lb]{\smash{{{\SetFigFont{8}{14.4}{\rmdefault}{\mddefault}{\updefault}$i$}}}}}
\put(4800,2637){\makebox(0,0)[lb]{\smash{{{\SetFigFont{8}{14.4}{\rmdefault}{\mddefault}{\updefault}$\bi$}}}}}
\put(4800,2037){\makebox(0,0)[lb]{\smash{{{\SetFigFont{8}{14.4}{\rmdefault}{\mddefault}{\updefault}$i$}}}}}
\put(4800,1437){\makebox(0,0)[lb]{\smash{{{\SetFigFont{8}{14.4}{\rmdefault}{\mddefault}{\updefault}$\bi$}}}}}
\put(4800,837){\makebox(0,0)[lb]{\smash{{{\SetFigFont{8}{14.4}{\rmdefault}{\mddefault}{\updefault}$i$}}}}}
\put(4800,237){\makebox(0,0)[lb]{\smash{{{\SetFigFont{8}{14.4}{\rmdefault}{\mddefault}{\updefault}$\bi$}}}}}
\put(5400,5037){\makebox(0,0)[lb]{\smash{{{\SetFigFont{8}{14.4}{\rmdefault}{\mddefault}{\updefault}$\bj$}}}}}
\put(5400,4437){\makebox(0,0)[lb]{\smash{{{\SetFigFont{8}{14.4}{\rmdefault}{\mddefault}{\updefault}$j$}}}}}
\put(5400,3837){\makebox(0,0)[lb]{\smash{{{\SetFigFont{8}{14.4}{\rmdefault}{\mddefault}{\updefault}$\bj_1$}}}}}
\put(5400,3237){\makebox(0,0)[lb]{\smash{{{\SetFigFont{8}{14.4}{\rmdefault}{\mddefault}{\updefault}$j_1$}}}}}
\put(5400,2637){\makebox(0,0)[lb]{\smash{{{\SetFigFont{8}{14.4}{\rmdefault}{\mddefault}{\updefault}$\bj$}}}}}
\put(5400,2037){\makebox(0,0)[lb]{\smash{{{\SetFigFont{8}{14.4}{\rmdefault}{\mddefault}{\updefault}$j_1$}}}}}
\put(5400,1437){\makebox(0,0)[lb]{\smash{{{\SetFigFont{8}{14.4}{\rmdefault}{\mddefault}{\updefault}$\bj_1$}}}}}
\put(5400,837){\makebox(0,0)[lb]{\smash{{{\SetFigFont{8}{14.4}{\rmdefault}{\mddefault}{\updefault}$j$}}}}}
\put(5400,237){\makebox(0,0)[lb]{\smash{{{\SetFigFont{8}{14.4}{\rmdefault}{\mddefault}{\updefault}$\bj$}}}}}
\put(6000,5037){\makebox(0,0)[lb]{\smash{{{\SetFigFont{8}{14.4}{\rmdefault}{\mddefault}{\updefault}$\bi$}}}}}
\put(6000,4437){\makebox(0,0)[lb]{\smash{{{\SetFigFont{8}{14.4}{\rmdefault}{\mddefault}{\updefault}$i$}}}}}
\put(6000,3837){\makebox(0,0)[lb]{\smash{{{\SetFigFont{8}{14.4}{\rmdefault}{\mddefault}{\updefault}$\bi$}}}}}
\put(6000,3237){\makebox(0,0)[lb]{\smash{{{\SetFigFont{8}{14.4}{\rmdefault}{\mddefault}{\updefault}$i$}}}}}
\put(6000,2637){\makebox(0,0)[lb]{\smash{{{\SetFigFont{8}{14.4}{\rmdefault}{\mddefault}{\updefault}$\bi$}}}}}
\put(6000,2037){\makebox(0,0)[lb]{\smash{{{\SetFigFont{8}{14.4}{\rmdefault}{\mddefault}{\updefault}$i$}}}}}
\put(6000,1437){\makebox(0,0)[lb]{\smash{{{\SetFigFont{8}{14.4}{\rmdefault}{\mddefault}{\updefault}$\bi$}}}}}
\put(6000,837){\makebox(0,0)[lb]{\smash{{{\SetFigFont{8}{14.4}{\rmdefault}{\mddefault}{\updefault}$i$}}}}}
\put(6000,237){\makebox(0,0)[lb]{\smash{{{\SetFigFont{8}{14.4}{\rmdefault}{\mddefault}{\updefault}$\bi$}}}}}
\put(6600,5037){\makebox(0,0)[lb]{\smash{{{\SetFigFont{8}{14.4}{\rmdefault}{\mddefault}{\updefault}$\bj_1$}}}}}
\put(6600,4437){\makebox(0,0)[lb]{\smash{{{\SetFigFont{8}{14.4}{\rmdefault}{\mddefault}{\updefault}$j_1$}}}}}
\put(6600,3837){\makebox(0,0)[lb]{\smash{{{\SetFigFont{8}{14.4}{\rmdefault}{\mddefault}{\updefault}$\bj$}}}}}
\put(6600,3237){\makebox(0,0)[lb]{\smash{{{\SetFigFont{8}{14.4}{\rmdefault}{\mddefault}{\updefault}$j$}}}}}
\put(6600,2637){\makebox(0,0)[lb]{\smash{{{\SetFigFont{8}{14.4}{\rmdefault}{\mddefault}{\updefault}$\bj_1$}}}}}
\put(6600,2037){\makebox(0,0)[lb]{\smash{{{\SetFigFont{8}{14.4}{\rmdefault}{\mddefault}{\updefault}$j$}}}}}
\put(6600,1437){\makebox(0,0)[lb]{\smash{{{\SetFigFont{8}{14.4}{\rmdefault}{\mddefault}{\updefault}$\bj$}}}}}
\put(6600,837){\makebox(0,0)[lb]{\smash{{{\SetFigFont{8}{14.4}{\rmdefault}{\mddefault}{\updefault}$j_1$}}}}}
\put(6600,237){\makebox(0,0)[lb]{\smash{{{\SetFigFont{8}{14.4}{\rmdefault}{\mddefault}{\updefault}$\bj_1$}}}}}

\put(0,2712){\makebox(0,0)[lb]{\smash{{{\SetFigFont{12}{14.4}{\rmdefault}{\mddefault}{\updefault}Figure 2}}}}}

\end{picture}
}

There are $(n+1)(m-n+1)$ degenerate ground states of the vertex model
associated with this lattice (by a ground state, 
we mean a configuration that contributes maximally
to the partition function sum).  
The values of the
spin variables in different 
ground states are shown in Figure 2, in which we use the notation
$j_1=m+n-(2i+j)$, $\bi=n-i$, $\bj=m-j$, $\bj_1=m-j_1$.
There is a different ground state corresponding to each choice of the
pair of integers $0\leq i\leq n$, $0\leq j\leq m$ that obey the requirement
$n\leq i+j\leq m$.
We 
label such a ground state by the integers $0\leq a \leq m-n$ and $0\leq
b \leq n$ defined by $a=m-i-j$ and $b=n-i$.  We are interested in the
infinite-volume limit of this lattice taken as follows: instead of
having periodic boundary conditions, we fix the spins at the
boundaries of
the lattice to be equal to one of these ground states.  We label the
partition function by $Z_{a,b}^M$, and consider this
lattice in the limit $M\ra \infty$.

The infinite lattice so defined may be split into four quadrants.  The
associated corner transfer matrices are labelled $A_{NW}(\z),
A_{SW}(\z), A_{SE}(\z)$ and $A_{NE}(\z)$.  A {\it path} is defined as
$\ket{p}=p(s), (s>0)$, where $p(s)$ is the value of the spin variable
on the {\it s}-th line (our convention is that $s$ increases from east to
west, and from south to north).  $A_{NW}(\z)$ acts on the path space
$P_{a,b}$ defined by 
\bea
P_{a,b}&=& \{\ket{p}\, |\; p(s)=\bar{p}(s;a,b),\ws
s>>0\},\ws\hb{where},\nn
\\[3mm]
\bar{p}(s;a,b)&=&\cases{n-b &if $s$ is odd;\cr m-n-a+b&if
  $s\equiv 2\bmod 4$;\cr a+b&if $s \equiv 4\bmod 4$.\cr} \nn\ena
We have $A_{NW}(\z):P_{a,b}\ra P_{a,b}$,
$A_{SW}(\z):P_{a,b}\ra P_{m-n-a,n-b}$, 
$A_{SE}(\z):P_{m-n-a,n-b}\ra
P_{m-n-a,n-b}$, and  $A_{NE}(\z):P_{m-n-a,n-b}\ra P_{a,b}$.
In fact, using 
the crossing symmetry of the \R-matrix, we can write 
\be A_{SW}(\z)=S
A_{NW}(-q^{-1}\z^{-1}),\quad  A_{SE}(\z)=S A_{NW}(\z) S,\quad
A_{NE}(\z)= A_{NW}(-q^{-1} \z^{-1})S.\nn\ee
Here $S:P_{a,b}\ra P_{m-n-a,n-b}$ is the operator defined by
\be
p(s)\to \cases{
n-p(s), & if $s$ is odd;\cr
m-p(s), &if $s$ is even.\cr}
\nn\ee

Baxter's arguement \cite{Bax82} then implies that we have $A_{NW}(\z)
=(\hb{const})\, \z^{-H_{CTM}}$, where $H_{CTM}$ is independent of $\z$ and
has a non-negative integer spectrum. 
The partition function in the infinite lattice is 
\be
Z_{a,b}=\Tr_{P_{a,b}}(A_{NE}(\z) A_{SE}(\z) A_{SW}(\z) A_{NW}(\z))\propto
 \Tr_{P_{a,b}}(q^{2H_{CTM}}).\nn\ee

The next step is to understand the path space $P_{a,b}$ in terms of
the representation theory of $\uq$. The $\uq$ modules relevant to our
discussion are the level $\ell>0$ irreducible highest weight modules
$V(\l_r^{(\ell)})$. 
These are generated by the level $\ell$ highest weight vector
$v_{\l_r^{(\ell)}}$ $(\l_r^{(\ell)}=r\L_1 +(\ell-r) \L_0, \;0\leq r\leq \ell)$, which obeys
$e_1 v_{\l_r^{(\ell)}}=e_0 v_{\l_r^{(\ell)}}=f_1^{r+1}v_{\l_r^{(\ell)}}=f_0^{\ell-r+1}v_{\l_r^{(\ell)}}=0$.
We use a grading element $D$ which acts on
$V(\l_r^{(\ell)})$ as 
$D(f_{i_1} f_{i_2} \cdots  f_{i_N} v_{\l_r^{(\ell)}})=N(f_{i_1} f_{i_2} \cdots
f_{i_N} v_{\l_r^{(\ell)}})$. Labelling $\cH_{a,b}=V(\l_a^{(m-n)}) \ot
V(\l_b^{(n)})$, we conjecture that we can 
identify $P_{a,b}=\cH_{a,b}$ and $H_{CTM}=D$. 
We have a proof
of this identification in the $q\ra 0$ limit, but omit it
here for lack of space.

Now we come to the identification of half transfer matrices.
There are two half transfer matrices for our lattice, associated
with the insertion of half-infinite spin $n/2$ and spin $m/2$ lines
respectively (see Figures 3a and 3b). 

\setlength{\unitlength}{0.0006in}
\begingroup\makeatletter\ifx\SetFigFont\undefined%
\gdef\SetFigFont#1#2#3#4#5{%
  \reset@font\fontsize{#1}{#2pt}%
  \fontfamily{#3}\fontseries{#4}\fontshape{#5}%
  \selectfont}%
\fi\endgroup%
{\renewcommand{\dashlinestretch}{30}
\begin{picture}(5937,4500)(-800,-400)
\drawline(1425,861)(2625,861)
\drawline(1545.000,891.000)(1425.000,861.000)(1545.000,831.000)
\drawline(1425,2061)(2625,2061)
\drawline(1545.000,2091.000)(1425.000,2061.000)(1545.000,2031.000)
\dashline{60.000}(1425,1461)(2625,1461)
\drawline(1545.000,1491.000)(1425.000,1461.000)(1545.000,1431.000)
\dashline{60.000}(1425,2661)(2625,2661)
\drawline(1545.000,2691.000)(1425.000,2661.000)(1545.000,2631.000)
\drawline(4725,861)(5925,861)
\drawline(4845.000,891.000)(4725.000,861.000)(4845.000,831.000)
\drawline(4725,2061)(5925,2061)
\drawline(4845.000,2091.000)(4725.000,2061.000)(4845.000,2031.000)
\dashline{60.000}(4725,1461)(5925,1461)
\drawline(4845.000,1491.000)(4725.000,1461.000)(4845.000,1431.000)
\dashline{60.000}(4725,2661)(5925,2661)
\drawline(4845.000,2691.000)(4725.000,2661.000)(4845.000,2631.000)
\drawline(2025,3561)(2025,261)
\drawline(1995.000,381.000)(2025.000,261.000)(2055.000,381.000)
\dashline{60.000}(5325,3561)(5325,261)
\drawline(5295.000,381.000)(5325.000,261.000)(5355.000,381.000)
\put(1860,-50){\makebox(0,0)[lb]{\smash{{{\SetFigFont{12}{14.4}{\rmdefault}{\mddefault}{\updefault} $j$}}}}}
\put(5160,-50){\makebox(0,0)[lb]{\smash{{{\SetFigFont{12}{14.4}{\rmdefault}{\mddefault}{\updefault} $j$}}}}}
\put(4700,3300){\makebox(0,0)[lb]{\smash{{{\SetFigFont{12}{14.4}{\rmdefault}{\mddefault}{\updefault} .}}}}}
\put(4700,3100){\makebox(0,0)[lb]{\smash{{{\SetFigFont{12}{14.4}{\rmdefault}{\mddefault}{\updefault} .}}}}}
\put(4700,2900){\makebox(0,0)[lb]{\smash{{{\SetFigFont{12}{14.4}{\rmdefault}{\mddefault}{\updefault} .}}}}}
\put(5700,3300){\makebox(0,0)[lb]{\smash{{{\SetFigFont{12}{14.4}{\rmdefault}{\mddefault}{\updefault} .}}}}}
\put(5700,3100){\makebox(0,0)[lb]{\smash{{{\SetFigFont{12}{14.4}{\rmdefault}{\mddefault}{\updefault} .}}}}}
\put(5700,2900){\makebox(0,0)[lb]{\smash{{{\SetFigFont{12}{14.4}{\rmdefault}{\mddefault}{\updefault} .}}}}}
\put(1400,3300){\makebox(0,0)[lb]{\smash{{{\SetFigFont{12}{14.4}{\rmdefault}{\mddefault}{\updefault} .}}}}}
\put(1400,3100){\makebox(0,0)[lb]{\smash{{{\SetFigFont{12}{14.4}{\rmdefault}{\mddefault}{\updefault} .}}}}}
\put(1400,2900){\makebox(0,0)[lb]{\smash{{{\SetFigFont{12}{14.4}{\rmdefault}{\mddefault}{\updefault} .}}}}}
\put(2400,3300){\makebox(0,0)[lb]{\smash{{{\SetFigFont{12}{14.4}{\rmdefault}{\mddefault}{\updefault} .}}}}}
\put(2400,3100){\makebox(0,0)[lb]{\smash{{{\SetFigFont{12}{14.4}{\rmdefault}{\mddefault}{\updefault} .}}}}}
\put(2400,2900){\makebox(0,0)[lb]{\smash{{{\SetFigFont{12}{14.4}{\rmdefault}{\mddefault}{\updefault} .}}}}}
\put(-700,1911){\makebox(0,0)[lb]{\smash{{{\SetFigFont{12}{14.4}{\rmdefault}{\mddefault}{\updefault}Figures
\quad\quad    3a }}}}}
\put(4050,1911){\makebox(0,0)[lb]{\smash{{{\SetFigFont{12}{14.4}{\rmdefault}{\mddefault}{\updefault}  3b}}}}}
\end{picture}
}

\noindent We shall identify these
with certain components of the following intertwiners:
\[\begin{array}{lllll}
\phi^A(\z)&:&V(\l_a^{(m-n)})\ot V(\l_b^{(n)}) &\ra &
V(\l_a^{(m-n)})\ot V(\s(\l_b^{(n)}))
         \ot V^{(n)}_{\z},\nn\\[3mm]
\phi^B(\z)&:&V(\l_a^{(m-n)})\ot V(\l_b^{(n)}) 
&\ra& V(\s(\l_a^{(m-n)})) \ot V^{(m-n)}_{\z} \ot V(\l_b^{(n)})\nn \\
&& &\ra& V(\s(\l_a^{(m-n)}))\ot V(\s(\l_b^{(n)}))
         \ot V^{(m)}_{\z},
\end{array}\]
where $\s$ exchanges the fundamental weights
$\L_0\leftrightarrow \L_1$.
These intertwiners are in turn defined in terms of the more elementary
intertwiner 
\be \Phi^{(k,k+\ell)}(\z):V^{(k)}_{\z}\ot 
V(\l_r^{(\ell)}) \ra 
V(\s(\l_r^{(\ell)}))\ot V^{(k+\ell)}_{\z},\quad {k,\ell\geq 1}.\label{elint}\ee
We define $\Phi^{(0,\ell)}(\z)$ as the intertwiner $V(\l_r^{(\ell)})\ra
V(\s(\l_r^{(\ell)})) \ot V^{(\ell)}_{\z}$.
We define the components of $\Phi^{(k,k+\ell)}(\z)$ by
\be\Phi^{(k,k+\ell)}(u^{(k)}_i\ot v) =\sl_{j=0}^{k+\ell}
\Phi^{(k,k+\ell)}_{i,j} (v)\ot u^{(k+\ell)}_j; &\quad v\in
V(\l_r^{(\ell)}).
\nn\ee
The normalisation of $\Phi^{(k,k+\ell)}(\z)$ 
is fixed by
the requirement
\be \bra{\s(\l_r^{(\ell)})}\Phi^{(k,k+\ell)}_{0,\ell-r}(\z)\ket{\l_r^{(\ell)}}=1.\nn\ee
The intertwiner $\Phi^{(k,k+\ell)}(\z)$ is the generalisation to level
$\ell\ge 1$ of the intertwiner introduced by Nakayashiki \cite{Nak96};
its crystal limit was considered in \cite{Nak96b} and \cite{kaka97}
(see also \cite{MW97}). 
$\Phi^{(k,k+\ell)}(\z)$ has the
following properties: 
\bac 
 \xi^{-D} \Phi^{(k,k+\ell)}_{i,j}(\z)\, \xi^{D} &=&
 \Phi^{(k,k+\ell)}_{i,j}(\z/\xi)\label{props1}\\[2mm]
  g^{(k,k+\ell)} \sl_{j=0}^{k+\ell}\Phi^{(k,k+\ell)}_{i,j}(-q^{-1}
  \z)\Phi^{(k,k+\ell)}_{k-i',k+\ell-j}(\z)&=&\d_{i,i'},\label{commrelns}\ea
where
$g^{(k,k+\ell)}= { \qp4{q^{2k+2}} / \qp4{q^{2(k+\ell)+2}}}$.

If $v\ot v'\in V(\l_a^{(m-n)})\ot
V(\l_b^{(n)})$, then we define $\phi^A(\z)$ and $\phi^B(\z)$ by
\ba{lllll}
\phi^A(v\ot v')&=& \sl_{j=0}^{n}\phi^A_j(v\ot v')\ot u^{(n)}_j,\quad
\phi^A_j(v\ot v')&=&v\ot \Phi^{(0,n)}_jv',\label{def1}\\[3mm]
\phi^B(v\ot v')&=& \sl_{j=0}^{m} \phi^B_j(v\ot v')\ot u^{(m)}_j,\quad
\phi^B_j(v\ot v')&=&\sl_{j'=0}^{m-n} \Phi^{(0,m-n)}_{j'}v\ot
\Phi^{(m-n,m)}_{j',j}v'.\ea
Here, for clarity, we have suppressed the $\z$ dependence of all 
our intertwiners.
It is the components $\phi^A_{j}(\z),$
$(j=0,\cdots,n),$ and  $\phi^B_{j}(\z),$ $(j=0,\cdots,m),$
that we identify with the lattice insertions shown in Figures 3a and
3b respectively.

Now define $\cH=\sl_{a,b}\cH_{a,b}$ with $0\leq a
\leq m-n$ and $0\leq b\leq n$, and 
$\cF=\cH\ot \cH^*$. 
We shall
identify $\cF$ with the space on which our full transfer matrix
acts. Note that we can view $\cF$ as a linear map on $\cH$ via the
canonical identification $\cH\ot \cH^*\simeq \End(\cH)$. The dual
space $\cF^*$ is defined via $\bra{f} g\rangle=\hbox{\rm{Tr}}_{\cH}(f\circ g)$.
The transfer matrix itself is defined as ${\cal T}(\z)=T^B(\z) \circ
T^A(\z)$, where $T^A(\z)$ and $T^B(\z)$ represent the insertion of double-infinite vertical
spin $n/2$ and spin $m/2$ lines respectively. They are given by
\ba{lllll} 
     T^A(\z) &=& \sl_{j=0}^{n} T^A_j(\z),\quad  T^A_j(\z)&=& 
     g^{(0,n)} \, \phi^A_{j}(\z) \ot \phi^A_{n-j}(\z)^t,\label{Tdef1}\\
     T^B(\z) &=& \sl_{j=0}^{m}T^B_j(\z),\quad T^B_j(\z)&=&
     g^{(0,m)} \, \phi^B_{j}(\z) \ot \phi^B_{m-j}(\z)^t,
\ea
where $t$ denotes the transpose.
The motivation for the identification of the full space and for the
definition of the transfer matrices is essentially the same as 
that described in \cite{JM}.

A vacuum state is a maximal eigenstate of $\cT(\z)$. There are
$(n+1)(m-n+1)$ degenerate vacua, labelled as $\vac_{a,b}\in
\cH_{a,b}\ot \cH_{a,b}^*$, with $0\leq a\leq m-n, 0\leq b\leq n$.
We conjecture that, expressed as an element of $\End(\cH)$, the
vacuum $\vac_{a,b}$ and its dual $_{a,b}\dvac$ are given by
\be \vac_{a,b}=(\chi^{(m-n)}_{a} \chi^{(n)}_b)^{-\hf} (-q)^D
 \pi_{a,b},\quad
_{a,b}\dvac=(\chi^{(m-n)}_{a} \chi^{(n)}_b)^{-\hf}\,
\pi_{a,b} \,(-q)^D.\nn\ee
Here, $\chi^{(\ell)}_r=\Tr_{V(\l_r^{(\ell)})}(q^{2D})$,
and $\pi_{a,b}$ is the
projector $\cH\ra \cH_{a,b}$. Using the definitions \mref{def1} and
\mref{Tdef1}, and
properties \mref{props1}, it is then simple to show that $\cT(\z)
\vac_{a,b}=\vac_{m-n-a,b}$. 

In order to consider other eigenstates of $\cT(\z)$ apart from the
vacua, we must first introduce some more intertwiners. On the
level $\ell$ irreducible highest weight modules, we have 
\bea
\Psi^{*\,s} (\xi):V^{(1)}_{\xi}\ot V(\l_r^{(\ell)}) 
\ra V(\l_{r+s}^{(\ell)}),\nn\\
\Phi^{s}(\xi): V(\l_r^{(\ell)}) \ra V(\l_{r+s}^{(\ell)}) \ot V^{(1)}_{\xi},
\nn\ena
where $s=\pm 1$ (or $\pm$ for short).
For $v\in V(\l_r^{(\ell)})$, we define components by
\bea \Psi^{*\,\pm}(u^{(1)}_{\ep}\ot v) &=&
 \Psi^{*\,\pm}_{\ep}(v),\quad (\ep=0,1),\nn\\
 \Phi^{\pm}(v)&=& \sl_{\ep=0}^{1} \Phi^{\pm}_{\ep}(v)\ot
 u^{(1)}_{\ep},\nn\ena
where we again suppress the $\xi$ dependence.
We fix the normalisation of these intertwiners to be
%
\bea
     \bra{\l_{r-1}^{(\ell)}} \Psi^{*\,-}_1(\xi)\ket{\l_r^{(\ell)}} &=&1, \quad 
     \bra{\l_{r+1}^{(\ell)}} \Psi^{*\,+}_0(\xi)\ket{\l_r^{(\ell)}} =1,\nn\\
     \bra{\l_{r-1}^{(\ell)}} \Phi_0^-(\xi)\ket{\l_r^{(\ell)}} &=&1, \quad 
     \bra{\l_{r+1}^{(\ell)}} \Phi_1^+(\xi)\ket{\l_r^{(\ell)}} =1.\nn\ena
They have
the following commutation relations with the intertwiner
$\Phi^{(k,l)}(\z)$ defined in equation \mref{elint}:
\bac
\Phi^{(k,l)}_{i,j}(\z) \Psi^{*\pm}_{\ep}(\xi)&=&
     \sl_{i',\epp}\Psi^{*\mp}(\xi)_{\epp} \Phi^{(k,l)}_{i',j}(\z)
      R^{(k,1)}(\z/\xi)^{i,\ep}_{i',\epp},\\[3mm]
\Phi^{(0,l)}_{j}(\z) \Psi^{*\pm}_{\ep}(\xi)&=&
     \Psi^{*\mp}(\xi)_{\ep} \Phi^{(0,l)}_{j}(\z) 
      \tau(\z/\xi),\\[3mm]
\Phi^{(0,l)}_{j}(\z)\Phi^{\pm}_{\ep}(\xi)&=&
\sl_{\ep',j'}R^{(1,l)}(\xi/\z)^{\ep',j'}_{\ep,j} 
\Phi^{\mp}_{\ep'}(\xi) \Phi^{(0,l)}_{j'}(\z),\label{pcomm}\ea
where
$ \tau(\z)=\z^{-1}{\qp4{q\z^2} \qp4{q^3\z^{-2}}\ov
                          \qp4{q\z^{-2}} \qp4{q^3\z^{2}}}$.

\vspace*{2mm}

Now we are in a position to define our intertwiners on the tensor product
space, and to complete the diagonalisation of our transfer matrix $\cT(\z)$.
The intertwiners we require are
\bea
\psi^{(\half)\,s}(\xi)&:&V^{(1)}_{\xi}\ot V(\l_a^{(m-n)}) \ot V(\l_b^{(n)}) \ra 
V(\l_{a+s}^{(m-n)}) \ot V(\l_b^{(n)})\quad \hb{and}\nn\\[2mm]
\psi^{(0)\,s,\bs}(\xi)&:&V(\l_a^{(m-n)}) \ot V(\l_b^{(n)}) \ra 
V(\l_{a+s}^{(m-n)}) \ot V^{(1)}_{\xi}\ot V(\l_b^{(n)}) \ra
V(\l_{a+s}^{(m-n)}) \ot V(\l_{b+\bs}^{(n)}),\nn\ena
where again $s,\bs=\pm 1$.
On the element $v\ot v'\in V(\l_a^{(m-n)}) \ot V(\l_b^{(n)})$, they are
given in terms of the above intertwiners by
\bea
\psi^{(\half)\,s}(\xi)(u^{(1)}_\ep \ot v\ot
v')&=&\psi^{(\half)\,s}_{\ep}(\xi)(v\ot v')
=\Bigl(\Psi^{*\,s}_{\ep}(\xi)v\Bigr)\ot v',
\nn\\[2mm]
\psi^{(0)\,s,\bs}(\xi)(v \ot v')
&=&\sl_{\ep=0}^1\Bigl(\Phi^{s}_{\ep}(\xi)v\Bigr)\ot 
\Bigl(\Psi^{*\,\bs}_{\ep}(\xi)v'\Bigr).\nn\ena

Using the commutation relations \mref{pcomm} and the unitarity property
of the \R-matrix, it is then straightforward to show that 
\bea
 T^A(\z) \psi^{(\hf)\,s}_{\ep}(\xi)\vac_{a,b}&=&
 \psi^{(\hf)\,s}_{\ep}(\xi)\vac_{a,n-b},\nn\\[2mm]
T^B(\z) \psi^{(\hf)\,s}_{\ep}(\xi)\vac_{a,b}&=&\tau(\z/\xi)
 \psi^{(\hf)\,-s}_{\ep}(\xi)\vac_{m-n-a,n-b},\nn\\[2mm]
 T^A(\z) \psi^{(0)\,s,\bs}(\xi)\vac_{a,b}&=&\tau(\z/\xi)
 \psi^{(0)\,s,-\bs}(\xi)\vac_{a,n-b},\nn\\[2mm]
T^B(\z) \psi^{(0)\,s,\bs}(\xi)\vac_{a,b}&=&
 \psi^{(0)\,-s,-\bs}(\xi)\vac_{m-n-a,n-b},\nn\ena
and hence that
\bea \cT(\z) \psi^{(\hf)\,s}_{\ep}(\xi)\vac_{a,b}&=&\tau(\z/\xi)
     \psi^{(\hf)\,-s}_{\ep}(\xi)\vac_{m-n-a,b},\nn\\
\cT(\z) \psi^{(0)\,s,\bs}(\xi)\vac_{a,b}&=&\tau(\z/\xi)
     \psi^{(0)\,-s,\bs}(\xi)\vac_{m-n-a,b}.\nn\ena
The two eigenstates $\psi^{(\hf)\,s}_{\ep}(\xi)\vac_{a,b}$
and $\psi^{(0)\,s,\bs}(\xi)\vac_{a,b}$ are spin $\hf$ and spin
$0$ states respectively. The spin $\hf$ excitation has a
$T^A(\z)$ eigenvalue of 1 and a $T^B(\z)$ eigenvalue of $\tau(\z/\xi)$.
The spin $0$ excitation has a
$T^A(\z)$ eigenvalue of $\tau(\z/\xi)$ and a $T^B(\z)$ eigenvalue of $1$.
This is the same pattern for the spectrum as that
observed in the Bethe Ansatz calculations of 
\cite{VMN94} for 
the $n=1,m=2$ case.
Multi-particle states arise
as compositions of $\psi^{(\hf)\,s_i}(\xi_i)_{\ep_i}$ and
$\psi^{(0)\,s_i',\bs_i'}(\xi_{i'})$ acting on $\vac_{a,b}$. The eigenvalue of
$\cT(\z)$ is then simply the product of all the $\tau(\z/\xi_i)$  and
 $\tau(\z/\xi_{i'})$ factors.

The S-matrix of the excitations in our model is given by the following
relations:
\bea
\psi^{(\hf)\,s_1}_{\ep_1}(\xi_1) \psi^{(\hf)\,s_2}_{\ep_2}(\xi_2) &=&
\sl_{\epp_1,\epp_2,s'_1,s'_2}
\psi^{(\hf)\,s'_2}_{\epp_2}(\xi_2) \psi^{(\hf)\,s'_1}_{\epp_1}(\xi_1)
\bar{R}^{(1,1)}(\xi)^{\epp_1,\epp_2}_{\ep_1,\ep_2}\; \wt{\l}{\mu}{\mu'}{\nu}{\xi}
\nn \\[6mm]
\psi^{(0)\,s_1,\bs_1}(\xi_1) \psi^{(0)\,s_2,\bs_2}(\xi_2) &=&
\sl_{s'_1,s'_2,\bs'_1,\bs'_2} 
 \psi^{(0)\,s'_2,\bs'_2}(\xi_2) \psi^{(0)\,s'_1,\bs'_1}(\xi_1) \;
 \wo{\l}{\mu}{\mu'}{\nu}{\xi}
\wt{\tilde{\l}}{\tilde{\mu}}{\tilde{\mu'}}{\tilde{\nu}}{\xi},\nn \\[6mm]
\psi^{(0)\,s_1,\bs_1}(\xi_1) \psi^{(\hf)\,s_2}_{\ep_2}(\xi_2) &=&
\sl_{s'_1,s'_2} \psi^{(\hf)\,s'_2}_{\ep_2}(\xi_2) \psi^{(0)\,s'_1,\bs_1}(\xi_1)
\;\wth{\l}{\mu}{\mu'}{\nu}{\xi},
\nn\ena
where $\xi=\xi_1/\xi_2$.
We use the notation that $\l,\mu,\mu',\nu$ are dominant integral weights
of level $m-n$, with $\l,\nu$ fixed and $\mu=\l+s_2 \,\bar{\rho}\,$,
$\mu'=\l+s'_1\,\bar{\rho}$\,, $\nu=\l+(s_1+s_2)\,\bar{\rho}\;$ 
(where $\bar{\rho}=\L_1-\L_0$).
The $s'_1,s'_2$ sum is over all $s'_1,s'_2={\pm 1}$ such that 
$\nu=\l+(s'_1+s'_2)\,\bar{\rho}$. The notation and summations for the
tildered quantities are the same, except that the weights 
$\tilde{\l},\tilde{\mu},\tilde{\mu'},\tilde{\nu}$
are of level $n$. The remaining
quantities are given for level $\ell$ weights by
\bea
\wo{\l^{(\ell)}_r}{\l^{(\ell)}_s}{\l^{(\ell)}_u}{\l^{(\ell)}_t}{\xi}&=&
{X(p^2\xi^{-2}) \ov X(p^2\xi^{2})}
\;\wec{\l^{(\ell)}_r}{\l^{(\ell)}_s}{\l^{(\ell)}_u}{\l^{(\ell)}_t}{\xi^2} 
\xi^{
\d_{t,s+1} -\,
\d_{r,u-1}},\nn \\[6mm]
\wt{\l^{(\ell)}_r}{\l^{(\ell)}_s}{\l^{(\ell)}_u}{\l^{(\ell)}_t}{\xi}&=&{X(\xi^{-2}) \ov X(\xi^{2})}
\;\wec{\l^{(\ell)}_r}{\l^{(\ell)}_s}{\l^{(\ell)}_u}{\l^{(\ell)}_t}{\xi^2} 
\xi^{
\d_{t,s+1} -\,
\d_{r,u-1}},\nn \\[6mm]
\wth{\l^{(\ell)}_r}{\l^{(\ell)}_s}{\l^{(\ell)}_u}{\l^{(\ell)}_t}{\xi}&=&
{X(p\xi^{-2}) \ov X(p\xi^{2})}
\;\wec{\l^{(\ell)}_r}{\l^{(\ell)}_s}{\l^{(\ell)}_u}{\l^{(\ell)}_t}{p^{-1} \xi^2} 
(-\xi q^{-(1+r)})^{\d_{t,s+1} -\,
\d_{r,u-1}}
\;q^{\d_{r,t}}.
\nn \ena
where $
X(z)={(z;p^2,q^4)_{\infty} (q^4 z;p^2,q^4)_{\infty} \ov 
(q^2 z;p^2,q^4)_{\infty}^2},\quad 
p=q^{\ell+2}, $
and $\overline{W}^1_{\ell}$ is the RSOS Boltzmann weight 
in the notation of equation (B.2) in \cite{JMO93}.

When $n=1$, $m=2$ (the alternating spin $\hf$, spin $1$ model), the
above commutation relations become
\bea
\psi^{(\hf)}_{\ep_1}(\xi_1) \psi^{(\hf)}_{\ep_2}(\xi_2)
&=& - \sl_{\epp_1,\epp_2} 
\psi^{(\hf)}_{\epp_2}(\xi_2) \psi^{(\hf)}_{\epp_1}(\xi_1)
R^{(1,1)}(\xi)^{\epp_1,\epp_2}_{\ep_1,\ep_2},\nn\\
\psi^{(0)}(\xi_1) \psi^{(0)}(\xi_2)
&=& - \psi^{(0)}(\xi_2) \psi^{(0)}(\xi_1),\nn\\[1mm]
\psi^{(0)}(\xi_1) \psi^{(\hf)}_{\ep_2}(\xi_2) &=&
\tau(\xi)
 \psi^{(\hf)}_{\ep_2}(\xi_2) \psi^{(0)}(\xi_1).\nn\ena
\noindent Here, we have suppressed
the appearance of $s_i$ and $\bs_i$, as in this case they  are determined
solely by the space on which the operators act. These relations
are consistent with the results of \cite{VMN94}.


Now we have reached the point where we can discuss
correlation functions
and the mixing of ground states.
A correlation function of the infinite-volume vertex
model we are discussing is by definition a ratio 
$Z_{a,b}^{\{i_N,\cdots,i_1\}}/Z_{a,b}$.
Here, $Z_{a,b}$ is the partition function 
described earlier, and 
$Z_{a,b}^{\{i_N,\cdots, i_1\}}$ is a similarly 
weighted sum over a restricted set of spin configurations. The
restriction is to include only those configurations for which
the spin variables on some specified $N$ 
edges have the fixed values $i_N,\cdots,i_1$.
Suppose that $N$ is odd,
and that the $N$ edges are vertical ones located successively
in one row. Then, our correlation function
will be given by the expression 
\bac &&F_{a,b}^{\{i_N,\cdots,i_1\}}=_{a,b}\!\!\dvac 
T^B_{i_{N}}(1) 
T^A_{i_{N-1}}(1) T^B_{i_{N-2}}(1)\cdots
T^A_{i_{2}}(1) T^B_{i_1}(1) \vac_{a',n-b},\\[2mm]
&&=(g^{(0,m)})^{N+1\ov 2} (g^{(0,n)})^{N-1\ov 2}
   (\chi^{(m-n)}_a\chi^{(n)}_{b})^{-1}
    \\&&\times \Tr_{\cH_{a,b}}\left(q^{2D}
\phi^B_{i_{N}}(-q^{-1})
\phi^A_{i_{N-1}}(-q^{-1})
\phi^B_{i_{N-2}}(-q^{-1}) \cdots 
\phi^A_{i_{2}}(-q^{-1})
\phi^B_{i_{1}}(-q^{-1}) \right.\\ &&\left.\quad \quad \quad \quad \quad \quad \quad \times
\phi^B_{m-i_{1}}(1) 
\phi^A_{n-i_{2}}(1) \cdots
\phi^B_{m-i_{N-2}}(1)
\phi^A_{n-i_{N-1}}(1)
\phi^B_{m-i_{N}}(1) \right)
\label{corrfn},\ea
where \hspace*{30mm}
$ a'=\cases{ m-n-a & for $N=1$ mod 4;\cr
                   a & for $N=3$ mod 4.\cr}\nn$
\\[4mm](See \cite{JM} for a detailed
explanation of this expression for the spin $\half$ case; 
the generalisation to our
case is straightforward).

Now let us consider the case when  $n=N=1$. 
Our formula \mref{corrfn} 
then gives us the $1$-point function which corresponds to a restricted sum
in which the local variable on the central edge of the
central vertical spin $m/2$ line is fixed (in an alternating spin
$\hf$, \  spin $m/2$ lattice).
Suppose instead that we had started out with an infinite lattice consisting
of just spin $m/2$ lines. The algebraic picture of such pure models is 
considered in \cite{idzal93}.
The question arises as to whether, and how, the $1$-point function 
in our alternating spin lattice might be related to the $1$-point
in the pure lattice. We can supply a straightforward answer to this question
using the algebraic analysis we have described. 

The argument is as follows:
when $n=1$, $m>1$, our half space is $\cH_{a,b}=V(\l_a^{(m-1)})\ot
V(\L_b)$ where $0\leq a \leq m-1$, $0\leq b\leq 1$. 
In this case, we
have the irreducible decomposition
\be V(\l_a^{(m-1)})\ot V(\L_b)
=\oplus_{c=0}^m V(\l_c^{(m)})\ot \Omega^{a,b}_{c},
\label{decomp}\ee
where the sum runs over the level $m$ weights $\l_c^{(m)}$, and
$\Omega^{a,b}_{c}$ is the space
of highest-weight states in 
$V(\l_a^{(m-n)})\ot V(\L_b)$. In a recent paper \cite{JS97},
Jimbo and Shiraishi constructed the action of the deformed
Virasoro
algebra on  $\Omega^{a,b}_{c}$ by making use of the operator
\be \psi^{(0)}(\z) :V(\l_a^{(m-1)})\ot V(\L_b) \ra 
V(\l_{a'}^{(m-1)})\ot V^{(1)}_{\z}\ot 
V(\L_{b}) \ra V(\l_{a'}^{(m-1)})\ot V(\L_{1-b}),\nn\ee
which we have defined above (again we suppress 
the $s$ and $\bs$ indices of $\psi^{(0)}(\z)$). One can prove that
\be\phi^B(\z)\psi^{(0)}(\xi)=\psi^{(0)}(\xi)\phi^B(\z). \nn\ee 
Hence, $\phi^B(\z)$ commutes with the action of the deformed
Virasoro algebra; it acts only on the space  $V(\l_c^{(m)})$
in \mref{decomp}.

Now consider the consequences for the $1$-point function discussed above,
(given by \mref{corrfn}
when $N=n=1$). The trace simply splits up into the trace over the
different level $m$ highest weight modules in \mref{decomp}.
Specifically, we have
\be F_{a,b}^{\{k\}}= {\sl_{c=0}^m 
F_c^{\{k\}}\chi^{(m)}_c \chi_{\Omega^{a,b}_{c}} \ov
\sl_{c=0}^m \chi^{(m)}_c \chi_{\Omega^{a,b}_c}
},
\label{mixing}\ee
where $F^{\{k\}}_c(\z)$, with  $(k=0,\cdots m)$,
is the corresponding $1$-point function for the pure
spin $m/2$ lattice with the ground state labelled by $c$, and
$\chi_{\Omega^{a,b}_c}=Tr_{\Omega^{a,b}_c}(q^{2D})$.
Thus, the $1$-point function for the alternating lattice 
with a particular
ground state is given as a linear combination of the $1$-point functions
associated with different ground states in the pure lattice;
there is a mixing of ground states.

Let us briefly discuss the relation between the above result
and Baxter's argument in \cite{Bax78}. Consider the simplest case;
in which the lattice consists of spin $\half$ and spin $1$ lines.
The idea of Baxter is to change the arrangement of the lines
without changing the configuration sums. This is possible because of
the Yang-Baxter equation for the Boltzmann weights. A configuration sum
consists of the contributions of the chosen ground state and
other states which are modified from it at finitely many edges.
Namely, it is a sum over diagrams consisting of those edges on which
the values of the local variables differ from the ground state values.
Usually, in such
an expansion, contributions of larger diagrams are of higher order in $q$.
Suppose this were true for the alternating lattice. In order to compute
the $1$-point function of a spin $1$ local variable on a given edge,
we could  move the spin $\hf$ lines far away from this edge.
Then, small diagrams of low order 
would only be associated with 
the spin $1$ lines near the chosen edge. 
This would imply that
the $1$-point functions of the alternating lattice should be the same as
the $1$-point functions of the pure spin $1$ lattice. This is in contradiction
with our results from the representation theory. However, the above
assumption about the size and the order of diagrams is not correct
in the alternating lattice. Consider a rectangular region consisting of
spin $1$ edges that is bordered by four spin $\hf$ lines.
Take the ground state configuration on this lattice
corresponding to be $(a,b)=(0,0)$. This ground state
consists of alternating values $0$ and $2$ on the spin $1$ edges in this
region, and $0$ and $1$ on the spin $\hf$ edge on its border. If we switch the
configuration of these variables (i.e., $0\leftrightarrow2$ and
$0\leftrightarrow1$), the order of the contribution is $q^2$ no matter how
large the region is. This explains the origin of mixing in 
the small $q$ expansion.

In fact we have performed such a small $q$ expansion for our
alternating
lattice, and 
calculated up to order $q^4$ all the 
different $1$-point functions corresponding to the left-hand side 
of \mref{mixing}
when $m=2$. The level two $1$-point functions on the right-hand side of the 
equation were calculated by Idzumi \cite{Idz94}. 
To this order, the results agree with the mixing formula 
and disagree with the pure $1$-point function.

Finally, suppose we wish to calculate
 a correlation function such
as \be_{a,b}\dvac T^A_{k}(\z) \vac_{a,n-b}
= g^{(0,n)}\,(\chi^{(m-n)}_a\chi^{(n)}_{b})^{-1}   \Tr_{\cH_{a,b}}\left(q^{2D}  \phi^A_{k }(-q^{-1}\z) \phi^A_{n-k}(\z) \right)\label{corrfn2}\ee
that only involves the insertion of spin $n/2$ lines into
our alternating lattice. From its definition \mref{def1},
$\phi^A$ acts just on the right-hand side of the
tensor product $H_{a,b}=V(\l_a^{(m-n)})\ot V(\l_b^{(n)})$,
and so \mref{corrfn2} is equal to
\be  g^{(0,n)}\,(\chi^{(n)}_{b})^{-1}   \Tr_{V(\l_b^{(n)})}
\left(q^{2D}  \Phi^{(0,n)}_{k }(-q^{-1}\z) \Phi^{(0,n)}_{n-k}(\z)
 \right).\nn
\ee
This is the $1$-point function associated with the pure spin $n/2$
model with a single ground state. There is no mixing of
ground states when we calculate correlation functions
involving only insertions of spin $n/2$ lines into our
alternating lattice. The physical argument that was 
previously wrong is now right. If we move
spin $m/2$ lines far away from our central fixed spin
$n/2$ lines, then diagrams that are of low order in $q$
in the small $q$ expansion are indeed small in size and
restricted to the pure spin $n/2$ region.

To summarise: we have constructed an algebraic picture
of infinite-volume 
alternating spin $n/2$, spin $m/2$ $(1\leq n<m)$ vertex
models. We have diagonalised the transfer matrix. The 
eigenstates are made up from spin $\hf$ and spin $0$ particle-like
excitations. We have computed the two-particle S-matrix elements
for these excitations.
A trace formula is given for local correlation functions. 
Specialising to $1=n<m$, we have expressed the $1$-point function
associated with a spin $m/2$ line as a linear combination of
the corresponding $1$-point functions in the pure spin $m/2$ model
with different ground states.
For the general alternating model $(1\leq n<m)$, we have shown that the $1$-point function associated
with a spin $n/2$ line is equal
to the corresponding $1$-point function in the
pure spin $n/2$ model, without mixing of ground states.

In this letter, we have presented in brief the main results of our
analysis of alternating spin vertex models. A more detailed explanation
will be published elsewhere.

\vspace*{7mm}

{\small
\noindent
We wish to thank Boris Feigin, Michio Jimbo and Masaki Kashiwara for 
discussions.
TM and RW thank Feodor Smirnov for his hospitality in Paris. 
RW, JH and SJK thank everyone in RIMS for their 
hospitality. RW
acknowledges funding from the RIMS/Isaac\,Newton\,Inst./JSPS/Roy.\,Soc. Exchange
Program and from an EPSRC Advanced Fellowship. 
SJK acknowledges the research grants by
Basic Science Research Institute Program,
Ministry of Education of Korea, BSRI-98-1414, and GARC-KOSEF
at Seoul National University.
}

\vspace*{5mm}
\noindent 
jhong@math.snu.ac.kr, sjkang@math.snu.ac.kr,\\
miwa@kurims.kyoto-u.ac.jp, r.a.weston@ma.hw.ac.uk

\newpage
\baselineskip=13pt  

\end{document}